# eBrainII: A 3 kW Realtime Custom 3D DRAM integrated ASIC implementation of a Biologically Plausible Model of a Human Scale Cortex

Dimitrios Stathis[*], Chirag Sudarshan[†], Yu Yang[*], Matthias Jung[‡], Syed Asad Mohamad Hasan Jafri[*],
Christian Weis[†], Ahmed Hemani[*], Anders Lansner[§], Norbert Wehn[†]

[*]KTH Royal Institute of Technology, Stockholm, Sweden. Email: stathis, yuyang2, jafri, hemani@kth.se

[†]University of Kaiserslautern, Germany. Email: sudarshan, weis, wehn@eit.uni-kl.de

[‡]Fraunhofer IESE, Germany. Email: matthias.jung@iese.fraunhofer.de

[§]Stockholm University and KTH Royal Institute of Technology, Sweden. Email: ala@kth.se

*Abstract*−The Artificial Neural Networks (ANNs) like CNN/DNN and LSTM are not biologically plausible and in spite of their initial success, they cannot attain the cognitive capabilities enabled by the dynamic hierarchical associative memory systems of biological brains. The biologically plausible spiking brain models, for e.g. cortex, basal ganglia and amygdala have a greater potential to achieve biological brain like cognitive capabilities. Bayesian Confidence Propagation Neural Network (BCPNN) is a biologically plausible spiking model of cortex. A human scale model of BCPNN in real time requires 162 TFlops/s, 50 TBs of synaptic weight storage to be accessed with a bandwidth of 200 TBs. The spiking bandwidth is relatively modest at 250 GBs/s. A hand optimized implementation of rodent scale BCPNN has been implemented on Tesla K80 GPUs require 3 kW, we extrapolate from that a human scale network will require 3 MW. These power numbers rule out such implementations for field deployment as advanced cognition engines in embedded systems. The key innovation that this paper reports is that it is *feasible* and *affordable* to implement real time BCPNN as a custom tiled ASIC in 28 nm technology with custom 3D DRAM - eBrain II - that consumes 3 kWs for human scale and 12 W for rodent scale cortex model. Such implementations eminently fulfill the demands for field deployment.

*Keywords*−Neuromorphic Computing, Biologically Plausible, BCPNN, Computer Architecture, Custom Memory, 3D-Memory, ASIC

## I. INTRODUCTION

A key limitation of today's successful ANN algorithms is that they cannot operate autonomously in a closed loop dynamic environment the way biological creatures can. The ANNs typically work in batch mode with single sensory modality; ANNs are principally restricted to classifying audio or visual patterns, one at a time but do not contribute to any further action based on classification result nor does the classification result influence future classification. In contrast, biological creatures decide and act based on active real-time, multi-modal perception and sensory fusion. Another limitation of current ANN models is that they have poor and ad-hoc mechanisms for multiple-timescale sequence processing and memory. The brain in contrast has evolved to select, process and memorize streaming information over many inherent time scales. This capability enables the amazing time coordinated motor performance of for instance a table tennis player. Additionally, it enables the efficient use of everyday memories with a spectrum of fast to slow encoding and decay dynamics, e.g., from volatile working memory, over hippocampal memory to long-term memory with life-long retention. Moreover, the use of error back-propagation over many levels is biologically implausible [1].

To achieve a brain-like cognitive capability, we need to adopt biologically plausible brain models with a proven function, e.g. efficient associative memory [2]. In this paper, we have adopted *Bayesian Confidence Propagation Neural Network* (BCPNN) as the candidate biologically plausible model. The BCPNN model in contrast to CNN has been mainly used for biologically realistic spiking neural network models of neocortex and basal ganglia. A modular cortical architecture in terms of minicolumns (MCUs) and hypercolumns (HCUs) was the basis for several models of cortical associative memory recall [3, 2, 4, 5], sequence learning and serial recall [6], olfactory perception [7], and decision making in basal ganglia [8, 9]. Recent simulations of human working memory have used the BCPNN synaptic plasticity rule in the context of biologically detailed models of word-list learning in prefrontal cortex [10]. In addition, BCPNN has been used in more abstract brain-like non-spiking neural network models [11, 12, 13].

This work builds on [14] and presents three novel contributions that separates it from the previous work. The key contributions are as follows:

1. We present a semi-formal method to co-explore the algorithm and architecture design space for a near super-computer class biologically plausible model of cortex.
2. Introduce a novel custom 3D-DRAM design with a BCPNN specific addressing scheme.
3. Benchmark against a GPU implementation and a best effort estimate implementation on SpiNNaker-2. We also analyze the results to show what makes eBrain-II non-incrementally more efficient compared to GPU and SpiNNaker-2.

## II. THE BCPNN MODEL

The BCPNN model features Hebbian-Bayesian synaptic plasticity for spiking neurons inspired by Bayesian statistics. Synaptic weights and intrinsic currents are adapted on-line upon arrival of single spikes, which initiates a cascade of temporally interacting memory traces representing plasticity-relevant changes in the signal transduction, e.g. NMDA receptor activation, Ca-influx, dendritic back-propagation, synaptic tagging, CaMKII activation, and protein synthesis [15]. BCPNN synaptic plasticity thus demonstrates spike-timing dependence, stable return to a set-point over long time scales, and high responsiveness despite this stability

We next introduce the computational model of BCPNN and justify the lazy evaluation model and derive its computational, storage and spiking bandwidth requirements. Finally, we discuss the parallelism in BCPNN. These details then become the basis for design and dimensioning of eBrain II in the next section.

### A. BCPNN Overview

The atomic neuronal unit in the BCPNN is *Mini-Column Unit* (**MCU**) that mimics 100 neurons in a cortical minicolumn whose outputs are strongly correlated. The dimensions we state next are for human scale BCPNN. Hyper Column Units (**HCUs**) encapsulate 100 MCUs that compete in a soft winner take all fashion. The BCPNN is a network of 2 million HCUs. Each HCU can receive on average 10,000 incoming pre-synaptic spikes/s coming from MCUs in other and the same HCU. Each HCU can produce 100 outgoing post-synaptic spikes/s. Each such outgoing post-synaptic spike is fanned out to 100 HCUs, effectively creating 10000 output spikes/s/HCU.

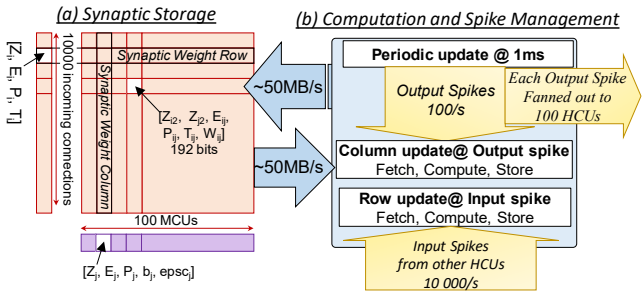

Figure 1. HCU structure and computations in a human scale BCPNN

#### 1. Synaptic Weights of BCPNN

In this section we explain the basis for the synaptic weights of BCPNN, see Fig. 1(a). The MCUs are modeled as leaky integrators and the input and output spikes, $S_i$ and $S_j$ respectively, are low pass filtered in three stages to generate traces called $Z_i$, $E_i$, $P_i$ and $Z_j$, $E_j$, $P_j$, see Fig. 2. Each of the traces has a different time constant, namely $\tau_z$, $\tau_e$ and $\tau_p$ as shown in 2. The index i represents the input to HCU at which the input spike arrives, there are 10 000 such inputs as shown in Fig. 1. Similarly, the index j represents the output of HCU that emits output spikes corresponding to the 100 MCUs in an HCU. The i traces are stored in an i-column vector and the j traces are stored in a j-row vector. The correlation of i and j traces for $Z$, $E$ and $P$ traces are stored in the $ij$-matrix. The $P$ traces, $P_i$, $P_j$ and $P_{ij}$, represent estimated probabilities and forms the basis for calculating the weight $w_{ij}$, which is also stored in the matrix. In addition, the bias $b_j$ of each MCU, representing its prior activation, is derived from the trace $p_j$. This bias plus the sum of weights from incoming spikes, determines the activation of each target MCU, the firing of which is then decided by a winner take all operation.

#### 2. Computation Model of BCPNN

Three kinds of operations happen in BCPNN, where the simulation time step is 1 ms. These operations are shown in Fig. 1(b).

*Periodic Update:* At every ms, the support vector is updated to decide which if any of the 100 MCUs has fired. On an average 100 output spikes are fired per second by each active HCU. Each of these spikes is fanned out to 100 HCUs as shown in Fig. 1.

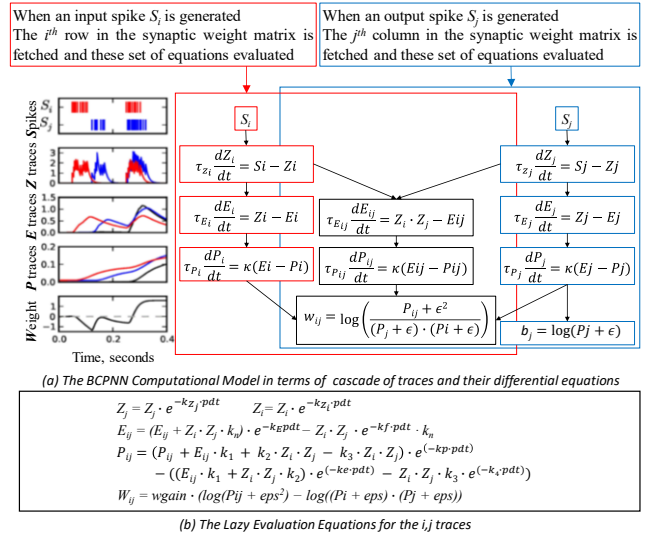

Figure 2. BCPNN computation model

*Column update:* If an output spike is fired as result of periodic update, at *ms* time boundaries, the column corresponding to the firing MCU is fetched from the synaptic storage, they are updated according to the equations shown in the blue box in Fig. 2(a) and stored back in the synaptic storage.

*Row update:* On an average 10000 spikes/s or 10 spikes/ms are received by each HCU. These spikes address specific rows in an HCU. The rows corresponding to each input spike is fetched from synaptic storage, updated according to equations shown in the red box in Fig. 2(a) and stored back.

*Lazy evaluation and Time stamping:* In principle, when there is no spike, the row and column vectors should be fetched and the traces decayed every *ms*. However, to avoid the costly storage access every *ms*, we time stamp the traces, $T_{ij}$, in Fig. 1, when we store them. When spikes arrive, the local time in *ms* is used to know the time elapsed since the last update and an integrated decay is applied. The cost-benefit analysis of such lazy evaluation has been reported in [16].

*The nature of BCPNN computation:* Many ANNs and scientific applications are modelled as matrix or vector multiplications. Inner loops of such applications are efficiently handled by GPUs. However, the kernel operation in BCPNN are far more complex as can be seen in Fig. 2(b). The customization of eBrainII synaptic storage and computational logic targets vectorizing such complex operations as we elaborate in section III and provides eBrain II with its superior efficiency.

#### 3. Bandwidth for Spike Propagation

The bandwidth for spike propagation in BCPNN is modest, 200 GBs/s. Fig. 3 shows the fields of a BCPNN spike and also the calculation for aggregate spike propagation bandwidth. Some fields are beyond the scope of this paper and are used during the structural plasticity phase of training. Two of three remain fields are obvious, Destination HCU and row. The third field Delay is used to emulate biological delay required for a spike to propagate from source HCU to destination HCU. Spike propagation in electronic systems are instantaneous by the *ms* standards of biology. For this reason, when an input spike arrives at an HCU, it is first stored in a *delay queue* for the delay *ms* duration before transferring it to *active queue*.

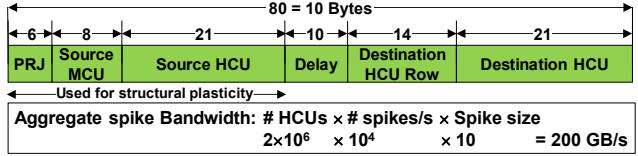

Figure 3. BCPNN Spike components and bandwidth.

## 4. Summary of BCPNN Requirements

We summarize the above discussion with the requirements of the lazy evaluation model in Table 1. The j-vector, being sufficiently small is stored locally excluded from the bandwidth requirement to synaptic storage.

**Table 1. Requirements for Lazy Evaluation model of human scale BCPNN**

|  | HCU | BCPNN |
|---|---|---|
| Computation | 81 MFlop/s | 162 TFlop/s |
| Storage | 25 MB | 50 TB |
| Bandwidth | 100 MB/s | 200 TB/s |
| Spike propagation | 100 kB/s | 200 GB/s |

### B. The Parallelism in BCPNN

BCPNN computation is embarrassingly parallel and deeply hierarchical as shown in Fig. 4. Each of the 2 million HCUs are independent threads with their own synaptic weight storage; there is no memory consistency problem.

Each HCU thread in turn has three concurrent threads for a) row computation, red box in Fig. 2 in response to the input spikes, b) column computation, blue box in Fig. 2, in response to the output spikes and c) the support computation that are periodically performed every *ms*. Though these three categories of threads are concurrent, each row and column update has to be atomic, the two cannot be interleaved because the two threads share the same storage; this is conceptually also shown by the overlap of red and blue boxes in Fig. 2. However, multiple row computations can be done in parallel as these computations do not share data. The same holds true for multiple column computations. The support computation, in contrast is independent and can be carried in parallel to both row and column computations.

Each row and column computation threads can themselves be split into embarrassingly parallel computations for each cell in row/column because the computation for each cell is independent of the computation in other cells.

Finally, each cell computation involves numeric solutions to a) the differential equations shown in Fig. 2 and b) integrated decay update. Both these computations for each of the three traces is a fairly complex and deep arithmetic flow graph that has rich potential for parallelism and some data dependencies that must be respected. This is shown as arithmetic level parallelism (ALP) in Fig.2, each box representing an additional floating point computational unit. This observation also applies to support computation.

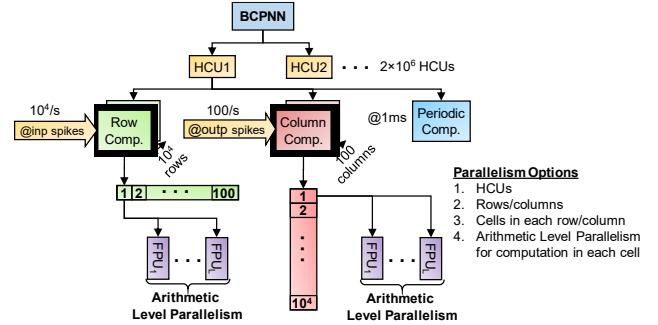

Figure 4. Parallelism options in BCPNN

## III. eBrain II: Design & Dimensioning

The design space for implementing BCPNN is spanned by the dimensions corresponding to the parallelism options discussed in section II.B. and shown in Fig. 4. The valid solutions in this space are the ones that fulfill the BCPNN requirements shown in Table 1 that the different technology options permit. In this section, we discuss the design and technology space for eBrain II and how did systematically made the design and dimensioning decisions.

### A. eBrain II's Hierarchical organization

The requirements for a human scale BCPNN as listed in Table 1 is obviously beyond a single chip solution for any technology available today. For this reason, eBrain II will be implemented as multiple chips called Brain Computation unit (**BCU**). Each BCU has resources to implement *N* HCUs, Fig. 5. There are two extreme options for how the resources for *N* HCUs can be organized.

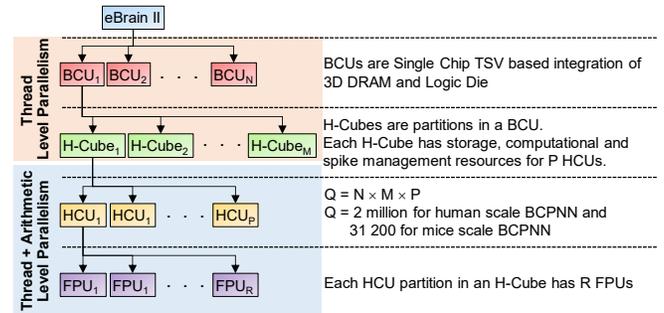

Figure 5. The hierarchical organization of eBrain

One extreme is to cluster the synaptic weights for all *N* HCUs in a single storage unit, **one vault**, accessed over a single shared channel for *N* non-deterministically concurrent, embarrassingly parallel HCU processes. The second extreme option is to have *N* vaults, each serving a single HCU. Since each HCU requires only 25 MBs, this would compromise the efficiency of the storage unit because the area of peripheral circuitry would overwhelm the area for bits. The optimal size of vault, as usual, is somewhere in between, where each vault integrates *P* HCUs. That gives sufficiently small DRAMs to have the benefit of high speed and low energy and acceptable overhead of the peripheral circuits. We call these vaults as ***H-Cubes***. In each H-Cube there are *P* parallel threads, each with its own *R* FPUs, control and scratchpad resources. Finally, each HCU thread, sequentially services row-computation, column-computation and periodic support-vector computation as atomic sub-threads, see sections II.A.2. Each of these three sub-threads is exposed to arithmetic level parallelism.

**B. The Design flow for dimensioning eBrain-II**

The design flow to dimension eBrain-II in terms of the sequence of steps, functional and technology constraints and their dependencies is shown in Fig. 6. The process starts by factoring in the real time constraint on the worst case number of input spikes that must be handled in each *ms*. This constraint is derived from the BCPNN specification that the *average* input spike rate is 10 spikes/*ms* and that they are poisson distributed. An additional constraint is imposed on average drop rate of one spike in a month. These forms the basis for the worst case *ms* constraint: eBrain-II should be able to deal with 36 spikes/*ms*. The direct implication of this is the size of the spike queue but also on the worst case bandwidth and computational load.

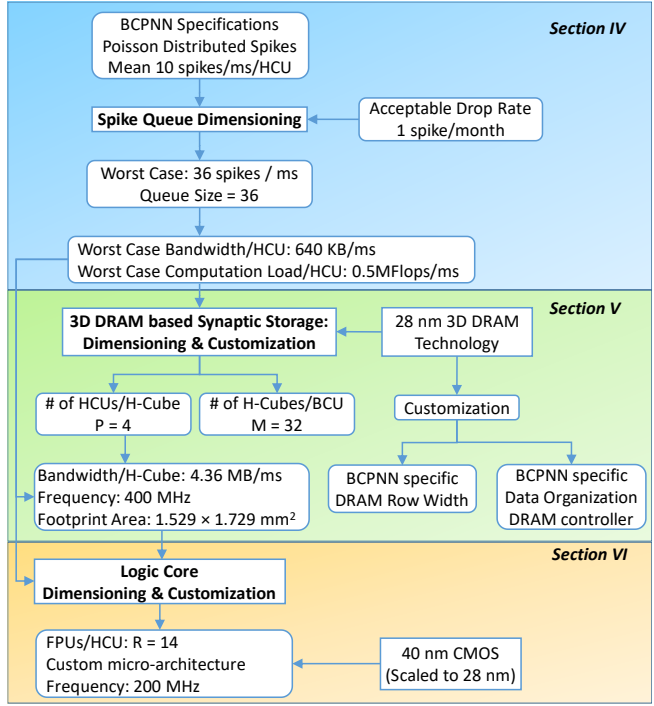

Figure 6. The Design flow for dimensioning eBrain-II

The second aspect of eBrain-II that is dimensioned is the synaptic storage system. In section V, we defend the choice of using custom 3D DRAM for the synaptic storage and how the *P* and *M* parameters were decided partly based on design space exploration and partly based on prevalent standards and pragmatics like improving yield. The channel width was customized for BCPNN weight matrix cell and the frequency adjusted to cover the worst case bandwidth. Additionally, the DRAM row/page size was customized to be of the same size as the BCPNN weight matrix row size and the data organization and memory controller are also customized for BCPNN. Both customizations improve utilization and lower the energy consumed by the synaptic storage sub-system.

The third and the last aspect of dimensioning concerns the logic that implements the BCPNN model. The logic core has to fulfill the worst case computational load (*0.5 MFLOP/ms*) and the area and frequency constraint imposed by the decision on *P* parameter while dimensioning the synaptic storage. Concretely, the logic for *P* HCUs cannot exceed the footprint of the synaptic storage for *P* HCUs, the two together form an H-Cube. The H-Cube logic should also have a frequency that is compatible with the DRAM frequency and also sufficiently fast to meet the worst case computational load. This in turn decides the number of FPUs, the *R* parameter. The micro-architecture for the logic core is customized for BCPNN.

In the next three sections, we elaborate the dimensioning of spike queue, synaptic storage and computational core for eBrain-II.

## IV. SPIKE QUEUE DIMENSIONING

To find an acceptable spike drop rate, we assumed that the spikes arrive at each HCUs with Poisson distribution; this is in accordance with the BCPNN specification. The mean value, $\lambda$, is *10* input spikes/ms; this is again as per BCPNN specifications: the average or mean number of input spikes per ms is 10. Next, we plotted the *complement* of cumulative poisson distribution (EQ1) as shown in Fig. 7.

$$Probablity\ of\ Dropping\ Spikes\ (x, \lambda) = 1 - \sum_{k=0}^{x-1} \frac{e^{-\lambda}\lambda^{x-1}}{k!} \quad EQ1$$

This plot shows the probability of having *x-or-more* i.e. *x+* spikes/ms. The left most value is *0+* spikes and that obviously has a probability of *1*, whereas the probability of having *10+* spikes is *~0.5* as expected. As can be seen, the probability reduces to near *0*, after *22+* spikes.

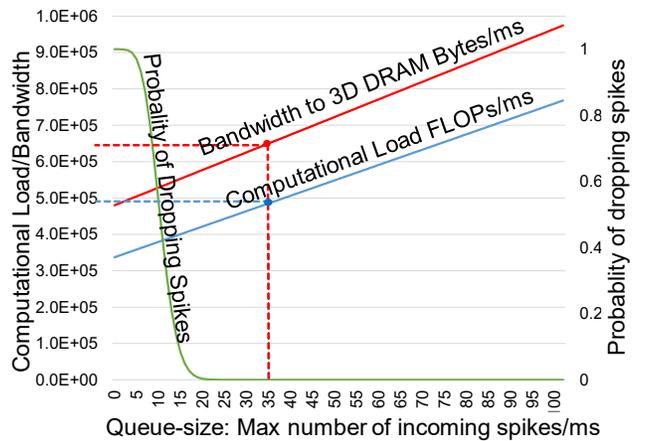

Figure 7. The probability of droping a spikes as a function of que size to deal with max number of spikes/ms and its impact on Computational Load and Bandwidth.

The *x+* values can also be interpreted as the queue-size.

For instance, if we have a queue-size of *10*, in each *ms*, there is a 50% chance that more than *10* spikes will arrive and spikes will be dropped. If instead we have a que size of *22*, there will be very few *ms(s)* in which more than *22* spikes will arrive. Based on these arguments, we selected the queue size to be *36* to reduce the probability of dropping spikes to near zero per *ms*; this implies a 30% probability of dropping spike once in a month.

An acceptable spike drop-rate decides the size of spike queue and this in turn provides constraint on computational load and the bandwidth to 3D DRAM. This is also shown in Fig. 7.

*Delay Queue Dimensioning:* As discussed in section II.A.3, there is a need to emulate biological delay. The queue dimensioning above is when the spike delay has been counted down to zero. We also need a spike for the delay to elapse that we call as delay queue. The delay queue dimension is derived from the active queue dimension by multiplying it by the average biological delay. For the human scale BCPNN, the average delay is 4 *ms*, which means that on an average a spike in the delay queue will persist for *4 ms* time slots before being moved to active queue. For this reason, we need the delay queue to be four times as large as the active queue.

### A. Constraints on Synaptic Storage Access and Computation

An important outcome of dimensioning the spike queue is that it sets the worst case bandwidth and computation load constraints. As discussed above, the worst case number of input spikes that eBrain II will deal with is 36. This implies that eBrain II should be dimensioned to do the following in this worst case *ms*: a) Do the periodic computation including storage access for read/write, b) Potentially do the column update in case periodic update results in an output spike. This also includes storage access and c) Do row updates corresponding to the 36 input-spikes, again including the storage access

An early design decision that we can easily make is that since periodic update happens irrespective of input or output spikes every *ms* and the data needed for it is modest 3.2 KB, it can be stored locally and not clog the bandwidth for the main synaptic weight matrix. Based on this, worst case bandwidth and computational load requirements are as follows: *640 MB/HCU/ms and 0.5 MFlop/ms.*

## V. SYNAPTIC WEIGHTS STORAGE SUB-SYSTEM

The synaptic storage and bandwidth requirements are the dominant requirements in Table 1 and in this section we justify the technology choice of custom 3D-DRAM, the dimensioning and architecture for the 3D DRAM and finally how did we customize the data organization and memory control to optimize the channel bandwidth.

### A. The Case for Custom 3D-DRAM

There are many emerging memory technologies with attractive attributes like non-volatility, low energy and high density etc., see [17]. However, DRAM still has the best combination of density, speed, endurance, write energy and manufacturing maturity including 3D integration. The maturity of the technology and its design methods provide us with accurate circuit level parametric design capability to customize the design and know its cost accurately.

Given the large storage and bandwidth requirements, 50 TB and 200 TBs/s, 3D integrated DRAM is an obvious choice. There are two competing standards for 3D integrated DRAM: HMC and HBM, [18, 19]. However, both of these options have a limited number of channels/links, maximum of four links in case of HMC and eight channels for HBM. Thus, targeting very large shared data of 1 to 2 GB over a high bandwidth shared channel/link. BCPNN is embarrassingly parallel at a much finer granularity; each HCU needs 25 MBs. To use HMC and HBM, would imply packing synaptic storage for 32 to 64 non-deterministically concurrent HCUs in a *large* memory over a *shared* channel. Shared channel for non-deterministically concurrent HCU processes could result in over-dimensioning the channel cope with peak traffic and underutilization when there is a slack in traffic. Finally, these memories cannot be custom dimensioned to match the dimensions of the BCPNN data structure to improve access efficiency. These considerations motivated us to settle for a custom 3D DRAM. Additionally, to fully exploit the high available bandwidth of TSVs, the custom 3D-DRAM is implemented using fine-grained rank-level partitioning and stacking architecture [20].

### B. The H-Cube dimension and architecture

The H-Cube dimension decides the number of HCUs (the *P* parameter in Fig. 5.) that share a channel to 3D DRAM vault that holds the synaptic storage. Smaller value of *P* implies higher architectural efficiency, because fewer HCUs would contend for the channel bandwidth. However, smaller value of *P* implies smaller DRAM partition which in turn lowers the *cell efficiency,* defined as the percentage of total area used by the storage cells to the total area including the peripheral circuits like row buffer, sense amplifier etc. The cell efficiency of commodity DRAMs is usually 45 - 55%. For this reason, we set the lower limit to be 40%. Thus, our exploration for 28 nm node resulted in 128 Mb memory tile per bank, see Fig. 7, as nearly optimum solution from the perspective of cell efficiency, vault area and energy consumption for this application [20]. Eight DRAM layers adds up to 1 Gb per vault that corresponds to synaptic storage for 5 HCUs, i.e. *P=5.* Since power of two simplifies addressing we settled for *P=4,* allowing us to also reserve some space for future extensions.

### C. Dimensioning TSV channel width and frequency

The HCU computations are in the granularity of one synaptic cell (192 bits). Hence, granularity of DRAM accesses is also set to 192 bits. This imposes the channel burst length (BL) times the data bus width (DQ) to be equal to 192 bits. The possible combinations of DQ and BL are (96, 2), (48, 4), (24, 8) etc. Higher values of DQ width results in a proportionally higher channel bandwidth, while adversely affecting the power dissipation and TSV area. Since, the aim of custom 3D DRAM is to have high number of channels, we choose the midrange DQ count of 48 and burst length of 4 as a good tradeoff between TSV area and DRAM power dissipation.

Finally, to address the bandwidth constraints for the worst-case *ms* the custom 3D-DRAM I/O frequency has to be tuned to a minimum of 400 MHz. With the presented dimensions the 3D-DRAM vault delivers an effective bandwidth of 4.35GB/s. This frequency not only achieves the bandwidth goal of 2.6GB/s, or 4×640 KB/ms as discussed in section IV, but also is in alignment with logic layer frequency of 200 MHz that we elaborate in section VI. In other words, the single cycle data transaction granularity (192 bits) of a HCU is equivalent to two DRAM cycle burst transactions.

To optimize the memory access, the row dimension was customized to be an exact multiple of synaptic weight matrix cell that requires 192 bits. Since there are 100 cells, see Fig. 1, the DRAM row size was dimensioned to be 100 × 192b. *The logical weight matrix row size is matched with the physical row size in the vault.*

### D. Dimensioning the BCU

Each BCU has one logic die and eight layers of stacked DRAM dies integrated with TSVs. The decision for 8 layers is based on what is possible in today's manufacturing technology. BCUs are divided into *M* equal sized partitions on each of the nine dies (8 DRAM + 1 Logic) and aligned to create *M* H-Cubes. Each H-Cube has a a memory vault that is private for the logic die underneath it as shown in Fig. 8. To reduce the overhead of peripheral circuits, a single tile per bank and single bank per die was chosen.

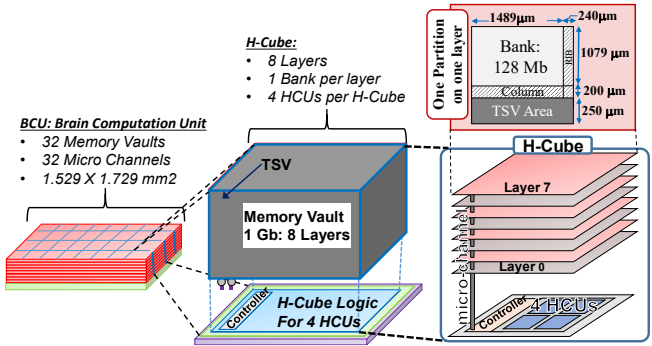

Figure 8. Organization of BCU in terms of M H-Cubes.

Given that each DRAM layer in a vault is 128 Mb, the present DRAM manufacturing norm suggests that we can either have 4Gb or 8Gb. In the interest of lower cost and higher yield we have selected 4 Gb per layer. For eight layers, this adds up to 32Gb, thus allowing 32 H-Cubes. The timing, current and area values of the presented custom 3D DRAM for the 28 nm process with the TSV pitch of 40 μm is extracted from DRAMSpec [21, 22] and the exact area dimension of each H-Cube is shown in Fig. 8.

### E. BCPNN specific data organisation

While the previous section focused on dimensioning of the 3D-DRAM for BCPNN, in this section we focus on the custom data organization for the BCPNN. This custom data organization is needed to optimize and achieve the desired bandwidth for the worst-case *ms*, as it was explained it section IV.

Row missed are the primary cause for the inadequate utilization of the maximum DRAM bandwidth that results in reduced performance, timing penalties and increased DRAM energy. The optimization of the DRAM physical address mapping is one of the approaches to reduce the number of row misses. There are a few prevailing address mapping schemes such as Bank Interleaving [23, 24], Permutation-Based Page Interleaving [25, 26], Bit-Reversal [27]. Apart from these general purpose address mapping schemes, a few memory controllers employ an address mapping that is tailored for a particular application, for example the Toggling Rate Analysis presented in [28, 29]. These Application Specific Memory Controllers (*ASMC*) [30] are designed from the overall system perspective and effectively utilize the application knowledge to largely improve the memory system bandwidth and reduces the DRAM energy in contrast to general purpose memory controllers. In our work, eBrain II, we have selected to implement such an ASMC.

A naive approach to store the big synaptic matrix of size 10000×100 BCPNN cells is direct mapping. This addressing will allow every row of the synaptic matrix to be mapped to a single DRAM row in the same bank. This simple technique is very effective for accessing an entire row of synaptic matrix, all 100 cells of that row are continuously available on the DRAM data bus without any DRAM row misses. In contrast, the column access of the synaptic matrix deteriorates the DRAM performance due to a huge number of DRAM row misses, i.e. one row miss per *i,j* cell access. This mapping imposes a high penalty on available DRAM bandwidth utilization and energy consumption. Identifying the low DRAM performance or inappropriateness of the prevailing address mapping schemes mentioned above for the BCPNN application, this work proposed a *novel application specific address mapping called Row-Merge*.

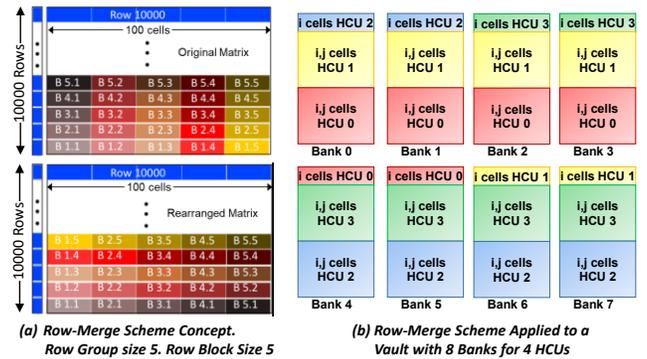

Figure 9. The proposed row-merge scheme to minimize DRAM row misses

The Row-Merge technique divides the 10000 rows of BCPNN synaptic matrix into equal groups, called as *row-groups*, each containing *X* rows. Every row is further divided into *X* equal *blocks*. All row-groups are rearranged to form new row-groups. The rearrangement is performed by placing the $y^{th}$ block of each row of an original row-group in the $y^{th}$ row of a new row-group. Fig. 9 (a) shows the rearrangement for *X = 5* where the block B1.3 in row 1, block 3 is moved to row 3, block 1. This is similar to block level interleaving in communication systems [31], however its use in DRAM is novel.

Row-Merge scheme maps the rearranged synaptic matrix row to a DRAM row, thus ensuring the BCPNN column access of $X_{i,j}$ cells per DRAM row hit because these previous column cells have now been rearranged as row cells. However, it will also result in $X$ DRAM row misses to access an entire synaptic matrix row. There are on an average 100 column updates and 10000 row updates per second. To find the best value of $X$, we should first identify the overall DRAM row misses per second due to these accesses and the $X$ value that amounts to a minimum row misses is selected. The read/write of a synaptic matrix row or a column results in $X$ or $100 \times (100/X)$ DRAM row misses respectively. Thus the overall DRAM row misses per second is formulated as shown by the equation in Fig. 10. The possible values of the $X$ are the factors of 100, where $X=1$ is nothing but direct address mapping scheme. The Fig. 10 shows the number row misses per second for each value of $X$. As can be seen $X=10$ is the best solution as gives the minimal number of row misses which is 5 times less compared to direct mapping.

$$Rowmiss = 10000 \cdot \left(X + \left(\frac{100}{X}\right)\right) \cdot 2$$

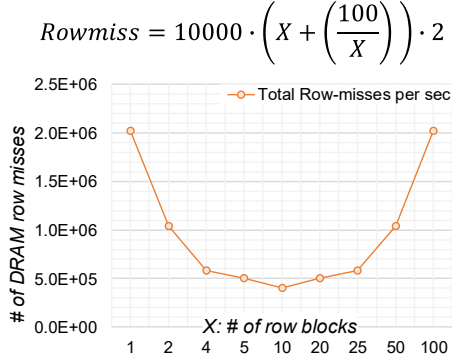

Figure 10. DRAM row misses as a function of number of blocks.

The rows of the rearranged synaptic matrix corresponding to each HCU of an H-cube are mapped to a DRAM row. They are row wise bank interleaved across 4 DRAM banks to further improve the DRAM, bandwidth due to bank level parallelism as shown in Fig. 9 (b). The above mentioned Row-Merge scheme and bank mapping is implemented in ASMC using address translation.

## VI. DIMENSIONING COMPUTATIONAL CORE

In this section, we explain how we designed and dimensioned the computational core to meet the worst case computational load constraint of 0.5 MFlop/ms/HCU derived in section IV. This worst case requirement is further constrained by the decisions made in the previous section V: 4 HCUs per H-Cube, area available is 2.582 mm$^2$ and the bandwidth available is 4.35 GB/s.

We divide the dimensioning of the computational core in terms of the local SRAM based storage, timing and computational resources required for the three types of updates in BCPNN as introduced in section II.A.2.

### A. The Local SRAM requirement

The BCPNN makes use of two different data-structures of different data size, namely synaptic columns and rows in Fig. 1. The synaptic columns are 100X bigger than rows in terms of data size. One row requires ~2.4 KBs which is moderate and it is reasonable to hold one or a few rows in scratchpad SRAM buffer(s). However, a BCPNN column will require an unreasonably large buffer to hold it. For this reason, column updates are split into 100 row sized updates. The periodic support-vector computation is relatively modest (11 KB) and is updated every *ms*. For this reason, the periodic update vector is not stored in DRAM but locally in an SRAM buffer.

### B. Time constraints for BCPNN computation

The worst-case *ms* load of the BCPNN updates is formalized in EQ2, expressed in terms of the three types of updates. The factor of 36 in EQ2 is the worst case number of incoming spikes per *ms*, as discussed previously. To explain in more detail, we give a description of each of the equation terms.

1. For 36 inputs spikes, 36 rows with 100 cells per row must be fetched, updated and written back to the 3D-DRAM
2. For 1 output spike, 1 column with 10 000 cells must be fetched, updated and written back to the 3D-DRAM
3. The periodic support vector computation must be performed; the data is stored locally in SRAM.

$$T_{worst\_case\_ms} = 36 \cdot T_{row} + T_{col} + T_{periodic} \leq 1ms \qquad \text{EQ2}$$

$$T_{row} = (2-k) \cdot (T_{DRAM} + T_{row\_comp}) + (k-1) \cdot MAX(T_{DRAM}, T_{row\_comp}) \qquad \text{EQ3}$$

$$T_{row\_comp} = T_{init} + \frac{100 \cdot T_{cell}}{\#FPU\_Sets} \qquad \text{EQ4}$$

EQ3 models the timing required for a row update ($T_{row}$) with or without ping pong buffers. With ping pong buffer, the DRAM latency is masked and without, the DRAM latency ($T_{DRAM}$) adds to the row computation latency ($T_{row\_comp}$). The factor k models the presence (k=2) or absence (k=1) of ping pong buffers. When k=2, the first term disappears and $T_{row}$ is decided by the second term that selects the dominant latency between DRAM latency and the row computation latency. When k=1, the second term disappears and the first term decides $T_{row}$ simply by addition of DRAM and computation latencies.

The computational load for row, column and periodic updates for the worst-case *ms* is approximately 15 million FLOPs. For the *k=1*, the non-ping-pong case, a significant portion of *ms* would be spent on DRAM access, making the task of fulfilling 15 MFlops more challenging. For this reason, we have opted for *k=2*, that is having ping pong buffers to hide the DRAM access latency.

With *k=2* design decision made, the first term in EQ3 disappears and $T_{row}$ will be decided by the dominant latency between $T_{DRAM}$ and $T_{row\_comp}$. The ideal design would balance them perfectly, so that one does not starve the other. Since the design decision on the size of DRAM macro per H-Cube is already made, see section V, the channel bandwidth and the speed at which it operates is decided, i.e. $T_{DRAM}$ is decided. One of the constraint on the DRAM macro design is to ensure that the $T_{DRAM}$ is small enough to meet the constraint specified by EQ2. This requires us then to design the datapath for row computation such that $T_{row\_comp}$ matches $T_{DRAM}$ as closely as possible. Since the row and column computations are very similar and dominate the overall computational load with periodic computation being trivially small in EQ2. We next explain how the computational resources were dimensioned to meet the timing constraints discussed in this section.

**C. Computational Resources**

Each HCU partition in an H-Cube exploits parallelism at two levels as discussed in section II.B. One is to update multiple cells in parallel and the other is to have arithmetic level parallelism while updating each cell. First we dimensioned the arithmetic level parallelism to update each cell and then we decided how many such parallel datapaths are needed to meet the timing constraints specified by EQ2. High-level synthesis is used to explore the design space for implementing $Z$, $E$ and $P$ updates while using different type of FPUs as shown in Fig. 11.

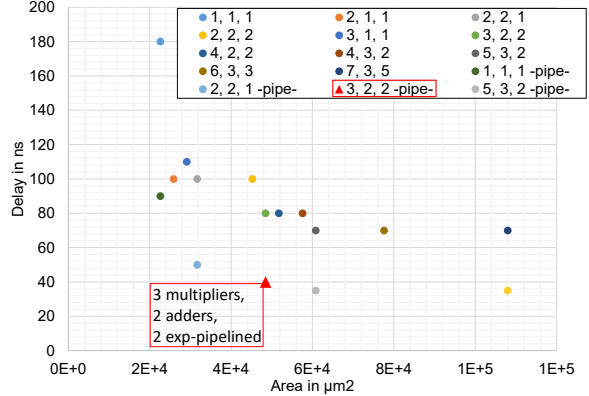

Figure 11. Design Space exploration for Arithmetic Level Parallelism in row/column cell computation.

Each point in the design space plot shows the number (#) and type of FPUs used as a triplet <#multiplier, #adder, #exp>. The log and divider are not included because one instance of them is needed for all cases. As expected and seen in the plot, in general, increasing area reduces the latency for one cell update, $T_{cell}$ in EQ4. The selected point is shown as red triangle with a boxed legend. This design point was selected because increasing area further has little impact on latency.

Once we have decided the arithmetic level parallelism for a row/cell computation and its latency ($T_{cell}$), we can decide how many cells must be updated in parallel to satisfy the EQ2. The factor #FPU_sets in EQ4 captures this cell-level-parallelism. Each *FPU_set* is the set of FPUs corresponding to the selected design point in Fig. 11. To satisfy EQ2, we need to update two cells in parallel requiring two *FPU_sets*.

We next explain, why we cannot increase the cell parallelism arbitrarily: There is a delay associated with filling the register files with data from scratchpad buffer (SRAM) and is represented by $T_{init}$ in EQ4. Since the scratchpad buffers, have single read and write ports, the register files of each datapath is filled in sequentially. This puts a limit on how many cells can be updated in parallel before $T_{init}$ starts to dominate.

**D. BCPNN Specific Infrastructure**

Besides the custom computational core and synaptic storage for BCPNN, eBrain-II also has BCPNN specific infrastructural resources. This includes, custom concurrent FSMs for managing computation, spike propagation and interaction with custom 3D DRAM based synaptic storage. These infrastructural operations degrade the GPU performance, as shown in section VIII.A. The organization of these infrastructural resources along with computational and local storage is shown in Fig. 12. We also have an aggressive power management policy to power down idle resources.

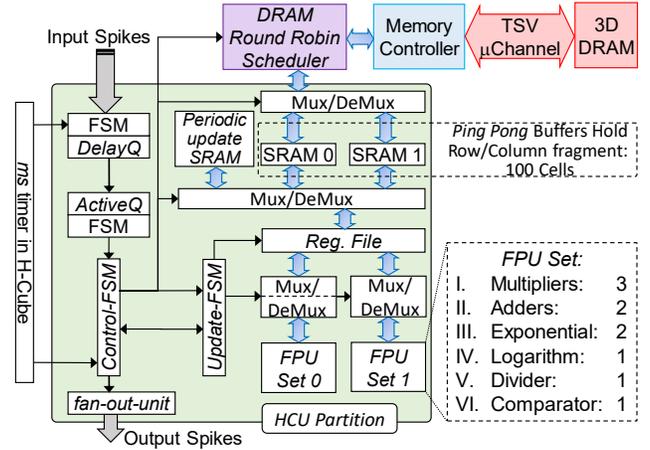

Figure 12. Detailed Schematic of an HCU Partition.

In each HCU partition we use a hierarchy of FSMs to control the update of the synaptic weights. For the row and column updates, the *control-FSM* fetches data into one of the *ping-pong* buffers by requesting the *DRAM-round-robin-scheduler*. For the periodic update, the data is stored locally in the *periodic-update-buffer* and not fetched from DRAM.

From the ping-pong or the periodic-update buffers, depending on the update, data is repeatedly transferred for two cells at a time into the *register-file* which is connected to the *FPU-sets* via *Mux/Demux logic*. The *FPU-sets* are themselves controlled by a separate FSM that takes care of the weight update. Once the update for two cells is done, the data is written back to the same buffer from which it was read. In parallel, data for the next row or column fragment is loaded into the other ping-pong buffer to pipeline the computation and DRAM access.

**E. Spike Propagation Network**

The spike propagation in eBrain II is also hierarchical reflecting the overall hierarchy of the architecture as shown in Fig. 5. Since an H-Cube is intended to cluster only a few HCUs *(P=4)* the spike propagation network and interface at H-Cube level is point to point with $P$ spike in and out ports.

At the BCU level, the input spike distribution network is a pipelined hierarchical binary tree, with pipeline stage at each node. This allows a regular log latency distribution and could be considered as a simple special-purpose hierarchical NOC. The output spike distribution network is similar but reversed in its direction of the data flow. Since the spike generation/arrival have *ms* latency, whereas these networks operate at *ns* granularity, the need for deep buffering and complex flow control that is typical of multi-core NOCs is absent. Instead, the buffered tree, serves both the purpose of buffering for long electrical wires and also serves as a spatially distributed queue. The generality and versatility of a full blown NOC would be an overkill for the intra-BCU spike propagation. The inter-BCU spike propagation details have not been refined yet but we argue that its impact on the overall cost will be trivially small. This is because a) the total spike propagation requirement is 3 orders less (Giga vs Tera) than the synaptic storage and computational requirements.

## VII. IMPLEMENTATION OF EBRAIN II

In this section, we elaborate the design methodology, the tool and the libraries that were used to implement eBrain II. In essence, we explain the experimental basis for the area, energy and performance numbers we claim. Table 2 presents the experimental setup. We then divide the presentation of the results into two sub-sections. The sub-section VII.A. details the major phases and sequence of the design flow and the sub-section VII.B. presents the detail analysis on area, energy and performance.

**Table 2. Experimental Setup**

| Technology Parameters | |
|---|---|
| Technology | 40 nm CMOS |
| Scaled Technology | 28 nm. The 40 nm node results were scaled using the conservative scaling factor of 0.71 based on [32]. |
| Syntheses | Cadence and Synopsys tools for High-level, Logic and Physical syntheses tools |
| Library | Synopsys DesignWare library for FPUs. |
| Simulation | SystemC TLM, RTL and gate-level HDL |
| Basis for Energy, Performance and Area Numbers Reported | |
| Logic | Simulation of post-layout data back annotated to gate-level models. |
| SRAM | SRAM Memory Compile Table |
| DRAM | DRAMSys circuit level models. [29] |
| Operating Parameters | |
| Voltage | 1.1 V 40 nm. 0.9 V for 28 nm |
| Frequency | 200 MHz. |
| Power Management | Power gating applied on per-HCU basis |
| | A wake up FSM in each HCU partition and the SRAM macro for Support Vector Periodic computation are not power gated |
| | Clock gating used during logic-synthesis |

### A. eBrain-II Design Steps

eBrain-II implementation was carried out in three major design phases as described below. A key objective of this sub-section is to give the readers a feel for the sequence of the effort involved in the design of eBrain-II.

#### 1. Phase I: Component Implementation and Dimensioning

In this phase we implemented the building blocks, characterized them and dimensioned the eBrain-II.

*Floating Point Units:* The floating point units (FPUs) down to physical level and each floating point unit was characterized with exhaustive gate level simulation back annotated with post-layout data to have sign-off quality assurance in the energy, area and latency numbers. The exponential unit was in the critical path and had 2 X longer logic depth compared to other FPUs. For this reason, we created a pipe-lined and non-pipe-lined units as the basis for design space exploration done by HLS tool. An important outcome of the FPU synthesis was the decision on frequency of operation. We used the largest logic depth of FPUs as the basis for the frequency of operation, i.e., 200 MHz.

*HLS Based DSE:* The BCPNN equations, the subset is shown in Fig. 2, were modeled in C for DSE using a public domain HLS tool [33]. The HLS library was populated with the synthesized FPUs. Note that HLS tool was used *only* for DSE, the actual RTL implementation was done manually in Phase II.

*Spike Queue Dimensioning:* The spike queue was dimensioned as explained in section IV.

*3D-DRAM Macro Dimensioning – the P Factor:* Using the DRAMsys design space exploration framework, design space for finding the optimal *P* factor, number of HCUs per H-Cube, was determined. As part of 3D DRAM macro size dimensioning, we also explored the data organization and how to optimize the channel bandwidth utilization. To achieve this, we created a stochastic model of BCPNN. This model did not have the actual computation but generated and absorbed spikes on a stochastic basis that mimicked the real BCPNN. This allowed us to generate a very large trace of DRAM access requests that was used to decide the data organization in DRAM and decide its dimensions.

*Datapath Dimensioning:* The datapath was dimensioned as explained in section VI. This essentially decides a) if ping-pong buffers are to be used or not, *k* factor in EQ3, b) the optimal FPU set from the HLS DSE and c) the number of parallel cells to be updated to satisfy EQ2.

#### 2. Phase II: RTL Design and Verifcation

In this phase we implemented the eBrain II hierarchy up to BCU level in RTL HDL and verified it.

*RTL Design:* For the H-Cube and HCU partitions for the specific dimensions that were decided in sections V and VI including the refinement of computational core as shown in Fig. 12. The hierarchical spike propagation network for integrating the HCUs in a BCU was also implemented.

*RTL Verification:* for the above RTL was carried out using a SystemVerilog based verification environment in a bottom up fashion against the golden BCPNN model in C++.

#### 3. Phase III: Logic and Physical Synthesis

In this phase, we performed the logic and physical synthesis of the RTL design down to physical level. The area, energy and latency results reported in this paper are based on post-layout sign off quality design data and exhaustive gate level simulation with back annotation of post layout data.

### B. Quantification of eBrain-II implementation results

In this section, we present the area, energy and performance numbers of eBrain II. The energy calculation for the ASIC implementation was done based on post-layout simulations. The design was synthesized down to physical level in 40nm technology and the numbers were scaled to 28 nm based on [32], i.e., area was scaled by 0.5 and power by 0.71. The SRAM energy numbers were based on memory compiler and the DRAM numbers were extracted using the DRAMSys design space exploration framework [29].

#### 1. eBrain-II Energy Numbers

The energy numbers are for an average spike rate of 10 input spikes/*ms* as per BCPNN specification. Fig. 13(a) shows the breakdown of energy in the form of a pie chart. As expected, the DRAM is the most dominant cost, in spite of a custom design and having minimal distance between DRAM and logic. The DRAM energy includes the core and I/O energy and that of memory controller.

On the logic die, the computation and SRAM consume the bulk of energy, whereas infrastructure that includes the spike queues, their controllers, spike propagation network and HCU master FSM only 3%. Note that the computation percentage includes FPUs, Register Files, Multiplexors, FSMs and *the wires*; FSMs, again as expected, take miniscule percentage and not shown as it would be invisible. *The energy reported is for dynamic and static power consumption. When the logic is idle, as stated before, we power them down*.

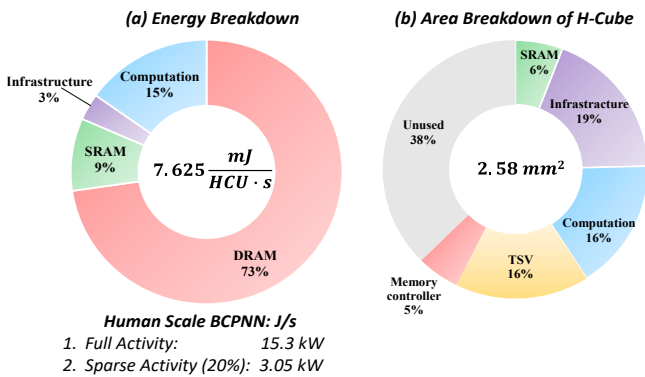

Figure 13. Energy and Area breakdown of BCPNN in 28 nm technology node.

We report two power numbers, one for full activity at human scale level with 2 million HCUs, all working. This is 15.3 kWs. This is artificial because it corresponds to every neuron firing at the same time. The activity in biological brain is very sparse. However, since the neuronal unit in BCPNN is MCU, which itself is an aggregation of 100 neurons, we report a number for 20% activity, which should be considered as a highly active cortex and this requires a modest 3 kWs.

#### 2. eBrain-II Area Numbers

The area breakdown is shown in Fig. 13(b). One can see that the infrastructure that had a small percentage of energy pie has a fairly large percentage of the area pie. This is because the spike queues are implemented using flops and there are quite many FSMs. However, the number of cycles spent in infrastructure is much smaller, typically at *ms* boundaries when spikes arrive, whereas once triggered, the computational logic works at every 5 *ns* cycle tick until each spike has been serviced.

Since the DRAM area and logic area overlap, the pie chart gives a breakdown of the logic die. Note that the DRAM macro for each H-Cube takes approximately twice as much area as the logic that is needed to do the computation and manage spikes. This is seen in Fig. 13(b) with the 38% *unused* part of the pie. To balance the footprint of synaptic storage and computational logic, we need a memory technology with higher density than what 8 layers of 3D DRAM integration can provide.

Each H-Cube has a footprint of 2.58 mm$^2$. Each BCU integrated 32 such H-Cubes making BCUs approximately 82.56 mm$^2$. If one factors in additional infrastructural blocks for clock, reset and power supply and pads, the area would approach 100 mm$^2$. For a human scale BCPNN, we would need 62.5K BCUs. To reduce this number, one could consider increasing the size of the die and/or integrating 16 such BCUs on an interposer to create super-BCUs.

#### 3. eBrain-II Performance Numbers

The eBrainII design achieves real time. For the worst-case *ms* when it needs to deal with 36 input spikes, all the necessary computation in EQ2 is achieved in 0.8 *ms* and for the average case of 10 spikes/*ms*, the computation is achieved in 0.2 ms. We did not use DVFS to fulfill worst-case with less logic area because we have ample area under 3D-DRAM that is not utilized, i.e., we are not area constrained. This is the reason why we opted for lowest voltage frequency operating point, highest level parallelism and power gating when idle as the most effective option.

### C. Rodent scale eBrain-II

The eBrain II design was customized and optimized for human scale BCPNN. The mice scale BCPNN has significantly lower requirements: 32K HCUs, 1200 rows and 70 columns. It is possible to map 16 rodent HCUs in each human scale H-Cube, even if such a mapping would underutilize some of the resources. With such a mapping we would need 62 BCUs, the mice scale BCPNN would run in real time and consume ~12 watts of power.

Table 3. Summary of the Experimental Results

| | | |
|---|---|---|
| Logic Area | 0.989 mm$^2$ | |
| ASMC Area | 0.135 mm$^2$ | |
| TSV Area | 0.423 mm$^2$ | |
| Memory Vault area | 2.582 mm$^2$ | |
| Energy consumption of eBrainII | Human | Mouse |
| | 3.05 KJ | 12 J |
| Effective bandwidth | Peak | Achieved Utilization |
| | 4.6875 GB/s | 4.3614 GB/s 93% |

| Dram Energy/bit | 7 pJ/bit |

## VIII. BCPNN ON OTHER PLATFORMS

In this section, we compare the BCPNN implementation on GPU and a best effort mapping of BCPNN to SpiNNaker-2 chip to eBrain II. We also briefly comment on why TrueNorth is not a suitable platform for BCPNN.

### A. BCPNN on GPU

BCPNN is mapped to NVIDIA Tesla K80 GPU card with dual GK210 cores, each with 12 GBs of DRAM. We used one of the cores and mapped 400 HCUs on it because the synaptic storage of 400 HCUs requires 10 GBs. We used the CUDA framework to map BCPNN that was composed in terms of four kernels: row update, column update, periodic update, queue management.

The *ms* tick in hardware is emulated by execution of a loop in CUDA. This *ms* loop includes the four kernels. Two of them, row, column are triggered by presence of spikes and the other two, the queue management and the periodic update are unconditionally executed.

When, as a result of periodic update, an MCU triggers and generates an output spike two actions happen. The first is that the column update kernel in source HCU is triggered and the second is that the output spike is fanned out to the destination HCUs as input spikes by copying these spikes to their respective delay queues. The queue management kernel decrements the biological delay in each iteration and when the delay reduces to zero, it triggers the row update kernels. This CUDA implementation was also verified against the golden C++ model of BCPNN.

To estimate power consumed by GK210, we used the in-built power measurement sensor to accurately know the power consumed and the CPU time was to know the time used and thereby energy. To find out how the energy was distributed among the major resource categories – storage, computation and infrastructure, we profiled CUDA BCPNN implementation on Nvidia Jetson TK1 for which we could use an accurate energy model [34]. The profiling data was fed to this energy model to know the percentage distribution of power among the resource categories. This percentage distribution was then applied to total energy measured on GK210.

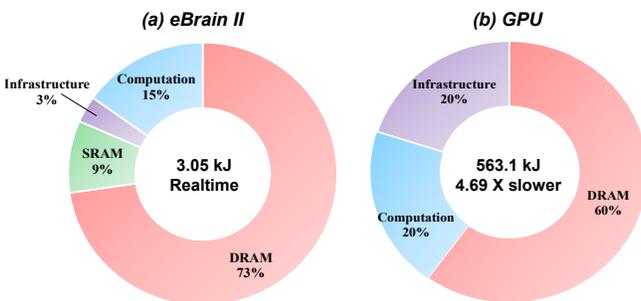

Figure 14. Human Scale BCPNN: eBrainII vs. GPU (GK210). 20% Sparse activity.

The key factors that degrades GPU's efficiency vis-à-vis eBrain-II are a) that the column access in GPU DRAM is poor compared to eBrain II's custom 3D DRAM, see section V, b) lack of custom concurrent FSMs that eBrain-II has to manage infrastructure, computation and DRAM access. All these functions in GPU are mapped to the same resources that also do the number crunching; this sequentializes the execution and underutilizes the resources. *As a consequence, the GPU implementation achieves 2645 kJs energy delay product that is 867 X worse off compared to eBrain II. Further, the GPU implementation delivers barely 23 effective GFLOP/s vs. the rated spec of 4365 GLFOP/s, i.e., only 5%.* The effective GFLOP/s are pure computation operations directly from the Lazy Evaluation model, excluding operations such as function calls, data rearrange, data copy, etc.

### B. BCPNN on SpiNNaker-2

We have also conceptually mapped BCPNN to SpiNNaker-2 system that is the next generation implementation of the SpiNNaker system. The SpiNNaker-2 architectural details are based on the early information available from [35] and email discussion with Prof. Stephen Furber at Univ. of Manchester, the principal architect of the system. The mapping is illustrated in Fig. 9.2. To maximize the utilization of the 144 cores and the 2GB available LPDDR4 DRAM, we propose mapping 72 HCUs to each SpiNNaker 2 chip. Each HCU's worst case *ms* load is divided into two cores to approximately balance the load among them. 72 HCUs also use 1.8 GB of DRAM.

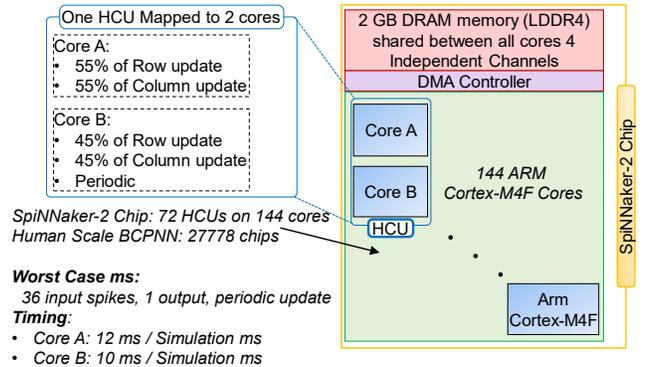

Figure 15. Conceptual Mapping of BCPNN to SpiNNaker-2

*Best Effort Estimation Method*

1. 250 MHz operation and 1.95 Instructions Per Cycle
2. BCPNN equations were compiled for Cortex M4F and simulated through OVP model [36].
3. Energy per instruction based on measured values for Cortex M4F [37] and scaled from 90 nm to 28 nm based on [38, 32].
4. The DRAM access latency is completely hidden by the DMA transfer in pipeline with the computation.
5. LPDDR4 bandwidth of 10 GB/s.
6. LPDDR4 energy assumed to be 5 pJ/bit based on [39]

Based on the above estimation method for a human scale BCPNN the total energy consumption for sparse activity (20%) is 220 kJ and it runs approximately 10X slower than real-time. This results in energy-delay product of 2200 kJ·s, i.e., 721X worse compared to eBrain II.

1. Each floating point operation in BCPNN equations results in on an average 7 instructions. This was found from the compilation results and simulation and is also consistent with the results reported in [37].
2. SpiNNaker-2, based on CortexM4F, like any processor needs fetch-decode-execute on a generic pipeline through 3-4 stages. eBrain in contras has a hard wired FSM that executes these floating operations in single cycle or in some cases two-three cycles for Exp, Log and Div operations.
3. CortexM4F has floating point operations but can only execute two of them in parallel (1.95 DMIPS/cycle). In contrast. eBrain II has 3 multipliers, 2 adders, 2 Exponentials, 1 Log and 1 Div FPUs for updating each cell.
4. CortexM4F implements Exp and Log functions in terms of primitive FP operations as library calls that further degrades the efficiency. eBrain II in contrast has custom hardware units for them.
5. GPU in comparison has massive parallelism for many cells because that is the granularity of parallelism and exploits the large number of FPUs available.

The energy and latency numbers reported does for the conceptual mapping of human scale BCPNN to SpiNNaker 2 did not include the Infrastructural operations including queue management and co-ordination between Spike arrival, DRAM access, and computation. We draw reader's attention to the fact that in case of GPU, the lack of custom concurrent FSMs that efficiently manages the infrastructural operations in eBrain II results in 60% degradation compared to the rated specs of the core. Further, we also ignore the Spike propagation cost. This is justified because in BCPNN it is proportionally insignificant compared to computation and synaptic weight storage. We have also given SpiNNaker 2 the benefit of unknown in assuming that LPDDR4 that lacks the customization in dimension and data organization of eBrainII will be as efficient in accessing rows and columns.

### C. BCPNN on TrueNorth and other platforms

TrueNorth [40] is a custom design to implement Neural Network models at fine granularity of neuronal models that are much simpler than the BCPNN neuronal unit – the MCU. The computational and storage units in TrueNorth are dimensioned to deal with simple leaky integrator and fire models of neurons, whereas the BCPNN, especially the lazy evaluation model involves far more complex operation and requires multi-level traces as described in section II.A.. For this reason, we did not find any reasonable way to even conceptually map BCPNN to TrueNorth and compare the results.

BCPNN, in past has been mapped to CRAY and IBM BlueGene super-computers, almost a decade ago. The energy delay product are obviously worse than what one can achieve with other platforms and we do not see that comparing eBrainII to these machines would bring any added scientific value. There are analog platforms like FACETs that are so different from eBrain II that it is hard to make a fair comparison and would require an unreasonably large effort.

## IX. CONCLUSION & FUTURE WORK

In this paper we have demonstrated that it is possible with modest engineering effort to do a custom ASIC design of a challenging biological plausible model of cortex like BCPNN stringent demands on real time to attain 162 TFlops/s and 200 TBs/s bandwidth to access 50 TBs of storage. The resulting design was shown to be three orders better than GPUs and SpiNNaker built on general purpose RISC micro-controllers. Custom logic and 3D DRAM storage with novel mapping enabled the power consumption is low enough for a rodent size biologically plausible cortex, 12W, to be deployed as cognition engine of embedded systems and the engineering effort is at least two orders less compared to ad hoc SOC designs of comparable complexity.

In future, we are planning eBrain- III that will have the following features:

1. It will operate on 22b integers rather than on single precision floating point operations
2. The BCPNN algorithm has been tweaked to eliminate the column updates and merge them with row updates
3. Use emerging non-volatile memories
4. The above changes will dramatically lower the computational, storage and bandwidth requirements to enable a human scale cortex to operate at ~500 W in real time
5. We will adopt synchoros VLSI design style based SiLago platform to automate transformation from untimed models in Matlab/Simulink like environment to timing and DRC clean GDSII. The synchoros SiLago platform, not only non-incrementally lowers the engineering cost, it can also lower the manufacturing cost by eliminating the mask engineering and DFT engineering costs, see [41] for more details.

## X. REFRENCES

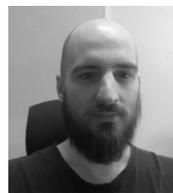

**Dimitrios Stathis** received his Diploma in Electr. & Comp. Engineering and a M.Sc. in Microelectronics and Computer Systems, both from the Democritus University of Thrace, Xanthi, Greece in 2013 and 2016 respectively. He received his second M.Sc. degree in Embedded Systems from KTH, Stockholm, Sweden in 2017. He is currently working towards the Ph.D. degree at the KTH, Stockholm, Sweden. His research interests include Computer Architectures and ASIC design for multimedia applications and Artificial Neural Networks.

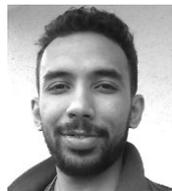

**Chirag Sudarshan** received his Master's (2017) degree in Electrical and Computer Engineering from the University of Kaiserslautern, Germany. He is working towards the PhD degree with the Microelectronic Systems Design Research Group, in the same university. His research interests are DRAM architecture, Process-in-Memory, DRAM controllers, and emerging memory technologies.

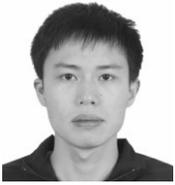 **Yu Yang** received his B.Sc degree in Automation from Sichuan University, Sichuan, China in 2011. In 2016 he received his M.Sc degree from Politecnico di Torino, Turin, Italy. He has been working towards the Ph.D at KTH Royal Institute of Technology, Stockholm, Sweden since 2016. His researchs focus on High-Level and System-Level Synthesis, EDA for HPC acceleration.

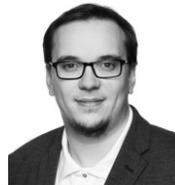 **Matthias Jung** received the Diploma and PhD degree in electrical engineering from the Technische Universität Kaiserslautern, Germany, in 2011 and 2017, respectively. From 2011 to 2017 he was a researcher Microelectronic Systems Design Research Group in the Electrical and Computer Engineering department of the Technische Universität Kaiserslautern. Since 2017 he is with the Embedded Systems Division of the Fraunhofer Institute for Experimental Software Engineering in Kaiserslautern as Senior Researcher and Project Manager. His research interest are embedded system hardware and software engineering, autonomous systems, memory systems, as well as, SystemC based virtual prototypes.

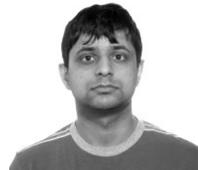 **Dr. Syed M. A. H. Jafri** is a senior system architect working at Ericsson. Before that he worked as a post-doc researcher at Royal Institute of Technology (KTH). He received his B.Sc. degree in 2005 from National University of Sciences and Technology in Rawalpindi, Pakistan. From 2005 to 2007 he was with Siemens, Pakistan. In 2009 he received MSc in system on chip design from Royal Institute of Technology (KTH), Sweden. He did PhD from University of Turku, Finland in 2015. He has 40+ peer reviewed journals/conference publications.

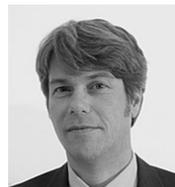 **Dr. Christian Weis** received the Ph.D. degree in electrical engineering from the TU Kaiserslautern, Germany, in 2014. From 1998 to 2009, he was with Siemens Semiconductor, Infineon Technologies AG and Qimonda AG, Munich, Germany, in DRAM design. In 2006, he was Design Team Leader for the 1Gb DDR3 DRAM, the first DDR3 volume product at Infineon/Qimonda. Since 2009, he has been with the Microelectronic System Design Research Group, TU Kaiserslautern, Germany. He holds several patents related to DRAM design, and published more than 60 papers. His current research interests include DRAM controllers, Near- & In-Memory processing, 3D-integrated DRAMs, and heterogeneous memory architectures.

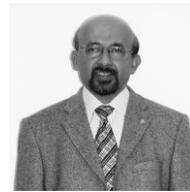 **Ahmed Hemani** is Professor in the Electronics Dept. at School of EECS, KTH, Stockholm, Sweden since 2006. His doctoral thesis on HLS was the basis for one of the first commercial HLS product from CADENCE. Later he has contributed to latency insensitive design style with research on GALS and GRLS. He also pioneered the concept of NOCs. His current research interests focus on massively parallel architecture and design methods for them to achieve ASIC comparable performance. He has introduced the concept of synchoricity and driven the development of the SiLago as an experimental synchoros VLSI design platform. He is applying synchoros VLSI design to artificial neural networks, biologically plausible models of brain and complex multi-media and telecom applications.

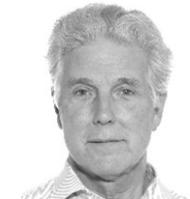 **Anders Lansner** is Professor in Computer science at KTH Royal Institute of Technology (KTH) in 1999, at Stockholm University since 2002. MSc in Chemical engineering from KTH in 1974, PhD in Computer science at KTH in 1986. Member of the Information Technology division of Royal Swedish Academy of Engineering Sciences since 2008. Main research interests include neuroinformatics, computational brain science, and brain-like computing with a focus on mathematical modelling and computer simulation of cortical associative memory, burst generation, and perception. Has co-authored more than 200 articles in these fields.

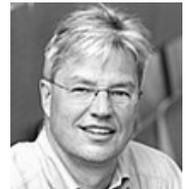 **Norbert Wehn** holds the chair for Microelectronic System Design in the department of Electrical Engineering and Information Technology at the University of Kaiserslautern. He received his Diploma and PhD from the TU Darmstadt in Germany. He has more than 300 publications in various fields of microelectronic system design and holds 20 patents. His special research interests are VLSI-architectures for mobile communication, forward error correction techniques, low-power techniques, advanced SoC and memory architectures, 3D integration, reliability issues in SoC, IoT and hardware accelerators for big data applications.